\documentstyle[twocolumn,seceq]{jpsj}

\edef\psfigRestoreAt{\catcode`@=\number\catcode`@\relax}
\catcode`\@=11\relax
\newwrite\@unused
\def\typeout#1{{\let\protect\string\immediate\write\@unused{#1}}}
\def\figurepath{./}

\def\@nnil{\@nil}
\def\@empty{}
\def\@psdonoop#1\@@#2#3{}
\def\@psdo#1:=#2\do#3{\edef\@psdotmp{#2}\ifx\@psdotmp\@empty \else
    \expandafter\@psdoloop#2,\@nil,\@nil\@@#1{#3}\fi}
\def\@psdoloop#1,#2,#3\@@#4#5{\def#4{#1}\ifx #4\@nnil \else
       #5\def#4{#2}\ifx #4\@nnil \else#5\@ipsdoloop #3\@@#4{#5}\fi\fi}
\def\@ipsdoloop#1,#2\@@#3#4{\def#3{#1}\ifx #3\@nnil 
       \let\@nextwhile=\@psdonoop \else
      #4\relax\let\@nextwhile=\@ipsdoloop\fi\@nextwhile#2\@@#3{#4}}
\def\@tpsdo#1:=#2\do#3{\xdef\@psdotmp{#2}\ifx\@psdotmp\@empty \else
    \@tpsdoloop#2\@nil\@nil\@@#1{#3}\fi}
\def\@tpsdoloop#1#2\@@#3#4{\def#3{#1}\ifx #3\@nnil 
       \let\@nextwhile=\@psdonoop \else
      #4\relax\let\@nextwhile=\@tpsdoloop\fi\@nextwhile#2\@@#3{#4}}
\newread\ps@stream
\newif\ifnot@eof       
\newif\if@noisy        
\newif\if@atend        
\newif\if@psfile       
{\catcode`\%=12\global\gdef\epsf@start{
\def\epsf@PS{PS}
\def\epsf@getbb#1{%
\openin\ps@stream=#1
\ifeof\ps@stream\typeout{Error, File #1 not found}\else
   {\not@eoftrue \chardef\other=12
    \def\do##1{\catcode`##1=\other}\dospecials \catcode`\ =10
    \loop
       \if@psfile
	  \read\ps@stream to \epsf@fileline
       \else{
	  \obeyspaces
          \read\ps@stream to \epsf@tmp\global\let\epsf@fileline\epsf@tmp}
       \fi
       \ifeof\ps@stream\not@eoffalse\else
       \if@psfile\else
       \expandafter\epsf@test\epsf@fileline:. \\%
       \fi
          \expandafter\epsf@aux\epsf@fileline:. \\%
       \fi
   \ifnot@eof\repeat
   }\closein\ps@stream\fi}%
\long\def\epsf@test#1#2#3:#4\\{\def\epsf@testit{#1#2}
			\ifx\epsf@testit\epsf@start\else
\typeout{Warning! File does not start with `\epsf@start'.  It may not be a PostScript file.}
			\fi
			\@psfiletrue} 
{\catcode`\%=12\global\let\epsf@percent=
\long\def\epsf@aux#1#2:#3\\{\ifx#1\epsf@percent
   \def\epsf@testit{#2}\ifx\epsf@testit\epsf@bblit
	\@atendfalse
        \epsf@atend #3 . \\%
	\if@atend	
	   \if@verbose{
		\typeout{psfig: found `(atend)'; continuing search}
	   }\fi
        \else
        \epsf@grab #3 . . . \\%
        \not@eoffalse
        \global\no@bbfalse
        \fi
   \fi\fi}%
\def\epsf@grab #1 #2 #3 #4 #5\\{%
   \global\def\epsf@llx{#1}\ifx\epsf@llx\empty
      \epsf@grab #2 #3 #4 #5 .\\\else
   \global\def\epsf@lly{#2}%
   \global\def\epsf@urx{#3}\global\def\epsf@ury{#4}\fi}%
\def\epsf@atendlit{(atend)} 
\def\epsf@atend #1 #2 #3\\{%
   \def\epsf@tmp{#1}\ifx\epsf@tmp\empty
      \epsf@atend #2 #3 .\\\else
   \ifx\epsf@tmp\epsf@atendlit\@atendtrue\fi\fi}

\def\psdraft{
	\def\@psdraft{0}
}
\def\psfull{
	\def\@psdraft{100}
}

\psfull

\newif\if@draftbox
\def\psnodraftbox{
	\@draftboxfalse
}
\@draftboxtrue

\newif\if@prologfile
\newif\if@postlogfile
\def\pssilent{
	\@noisyfalse
}
\def\psnoisy{
	\@noisytrue
}
\psnoisy
\newif\if@bbllx
\newif\if@bblly
\newif\if@bburx
\newif\if@bbury
\newif\if@height
\newif\if@width
\newif\if@rheight
\newif\if@rwidth
\newif\if@clip
\newif\if@verbose
\def\@p@@sclip#1{\@cliptrue}

\def\@p@@sfile#1{\def\@p@sfile{null}%
	        \openin1=#1
		\ifeof1\closein1%
		       \openin1=\figurepath#1
			\ifeof1\typeout{Error, File #1 not found}
			\else\closein1
			    \edef\@p@sfile{\figurepath#1}%
                        \fi%
		 \else\closein1%
		       \def\@p@sfile{#1}%
		 \fi}
\def\@p@@sfigure#1{\def\@p@sfile{null}%
	        \openin1=#1
		\ifeof1\closein1%
		       \openin1=\figurepath#1
			\ifeof1\typeout{Error, File #1 not found}
			\else\closein1
			    \def\@p@sfile{\figurepath#1}%
                        \fi%
		 \else\closein1%
		       \def\@p@sfile{#1}%
		 \fi}

\def\@p@@sbbllx#1{
		\@bbllxtrue
		\dimen100=#1
		\edef\@p@sbbllx{\number\dimen100}
}
\def\@p@@sbblly#1{
		\@bbllytrue
		\dimen100=#1
		\edef\@p@sbblly{\number\dimen100}
}
\def\@p@@sbburx#1{
		\@bburxtrue
		\dimen100=#1
		\edef\@p@sbburx{\number\dimen100}
}
\def\@p@@sbbury#1{
		\@bburytrue
		\dimen100=#1
		\edef\@p@sbbury{\number\dimen100}
}
\def\@p@@sheight#1{
		\@heighttrue
		\dimen100=#1
   		\edef\@p@sheight{\number\dimen100}
}
\def\@p@@swidth#1{
		\@widthtrue
		\dimen100=#1
		\edef\@p@swidth{\number\dimen100}
}
\def\@p@@srheight#1{
		\@rheighttrue
		\dimen100=#1
		\edef\@p@srheight{\number\dimen100}
}
\def\@p@@srwidth#1{
		\@rwidthtrue
		\dimen100=#1
		\edef\@p@srwidth{\number\dimen100}
}
\def\@p@@ssilent#1{ 
		\@verbosefalse
}
\def\@p@@sprolog#1{\@prologfiletrue\def\@prologfileval{#1}}
\def\@p@@spostlog#1{\@postlogfiletrue\def\@postlogfileval{#1}}
\def\@cs@name#1{\csname #1\endcsname}
\def\@setparms#1=#2,{\@cs@name{@p@@s#1}{#2}}
\def\ps@init@parms{
		\@bbllxfalse \@bbllyfalse
		\@bburxfalse \@bburyfalse
		\@heightfalse \@widthfalse
		\@rheightfalse \@rwidthfalse
		\def\@p@sbbllx{}\def\@p@sbblly{}
		\def\@p@sbburx{}\def\@p@sbbury{}
		\def\@p@sheight{}\def\@p@swidth{}
		\def\@p@srheight{}\def\@p@srwidth{}
		\def\@p@sfile{}
		\def\@p@scost{10}
		\def\@sc{}
		\@prologfilefalse
		\@postlogfilefalse
		\@clipfalse
		\if@noisy
			\@verbosetrue
		\else
			\@verbosefalse
		\fi
}
\def\parse@ps@parms#1{
	 	\@psdo\@psfiga:=#1\do
		   {\expandafter\@setparms\@psfiga,}}
\newif\ifno@bb
\def\bb@missing{
	\if@verbose{
		\typeout{psfig: searching \@p@sfile \space  for bounding box}
	}\fi
	\no@bbtrue
	\epsf@getbb{\@p@sfile}
        \ifno@bb \else \bb@cull\epsf@llx\epsf@lly\epsf@urx\epsf@ury\fi
}	
\def\bb@cull#1#2#3#4{
	\dimen100=#1 bp\edef\@p@sbbllx{\number\dimen100}
	\dimen100=#2 bp\edef\@p@sbblly{\number\dimen100}
	\dimen100=#3 bp\edef\@p@sbburx{\number\dimen100}
	\dimen100=#4 bp\edef\@p@sbbury{\number\dimen100}
	\no@bbfalse
}
\def\compute@bb{
		\no@bbfalse
		\if@bbllx \else \no@bbtrue \fi
		\if@bblly \else \no@bbtrue \fi
		\if@bburx \else \no@bbtrue \fi
		\if@bbury \else \no@bbtrue \fi
		\ifno@bb \bb@missing \fi
		\ifno@bb \typeout{FATAL ERROR: no bb supplied or found}
			\no-bb-error
		\fi
		\count203=\@p@sbburx
		\count204=\@p@sbbury
		\advance\count203 by -\@p@sbbllx
		\advance\count204 by -\@p@sbblly
		\edef\@bbw{\number\count203}
		\edef\@bbh{\number\count204}
}
\def\in@hundreds#1#2#3{\count240=#2 \count241=#3
		     \count100=\count240	
		     \divide\count100 by \count241
		     \count101=\count100
		     \multiply\count101 by \count241
		     \advance\count240 by -\count101
		     \multiply\count240 by 10
		     \count101=\count240	
		     \divide\count101 by \count241
		     \count102=\count101
		     \multiply\count102 by \count241
		     \advance\count240 by -\count102
		     \multiply\count240 by 10
		     \count102=\count240	
		     \divide\count102 by \count241
		     \count200=#1\count205=0
		     \count201=\count200
			\multiply\count201 by \count100
		 	\advance\count205 by \count201
		     \count201=\count200
			\divide\count201 by 10
			\multiply\count201 by \count101
			\advance\count205 by \count201
		     \count201=\count200
			\divide\count201 by 100
			\multiply\count201 by \count102
			\advance\count205 by \count201
		     \edef\@result{\number\count205}
}
\def\compute@wfromh{
		\in@hundreds{\@p@sheight}{\@bbw}{\@bbh}
		\edef\@p@swidth{\@result}
}
\def\compute@hfromw{
		\in@hundreds{\@p@swidth}{\@bbh}{\@bbw}
		\edef\@p@sheight{\@result}
}
\def\compute@handw{
		\if@height 
			\if@width
			\else
				\compute@wfromh
			\fi
		\else 
			\if@width
				\compute@hfromw
			\else
				\edef\@p@sheight{\@bbh}
				\edef\@p@swidth{\@bbw}
			\fi
		\fi
}
\def\compute@resv{
		\if@rheight \else \edef\@p@srheight{\@p@sheight} \fi
		\if@rwidth \else \edef\@p@srwidth{\@p@swidth} \fi
}
\def\compute@sizes{
	\compute@bb
	\compute@handw
	\compute@resv
}
\def\psfig#1{\vbox {
	\ps@init@parms
	\parse@ps@parms{#1}
	\compute@sizes
	\ifnum\@p@scost<\@psdraft{
		\if@verbose{
			\typeout{psfig: including \@p@sfile \space }
		}\fi
		\special{ps::[begin] 	\@p@swidth \space \@p@sheight \space
				\@p@sbbllx \space \@p@sbblly \space
				\@p@sbburx \space \@p@sbbury \space
				startTexFig \space }
		\if@clip{
			\if@verbose{
				\typeout{(clip)}
			}\fi
			\special{ps:: doclip \space }
		}\fi
		\if@prologfile
		    \special{ps: plotfile \@prologfileval \space } \fi
		\special{ps: plotfile \@p@sfile \space }
		\if@postlogfile
		    \special{ps: plotfile \@postlogfileval \space } \fi
		\special{ps::[end] endTexFig \space }
		\vbox to \@p@srheight true sp{
			\hbox to \@p@srwidth true sp{
				\hss
			}
		\vss
		}
	}\else{
		\if@draftbox{		
			\hbox{\fbox{\vbox to \@p@srheight true sp{
			\vss
			\hbox to \@p@srwidth true sp{ \hss \@p@sfile \hss }
			\vss
			}}}
		}\else{
			\vbox to \@p@srheight true sp{
			\vss
			\hbox to \@p@srwidth true sp{\hss}
			\vss
			}
		}\fi

	}\fi
}}
\def\psglobal{\typeout{psfig: PSGLOBAL is OBSOLETE; use psprint -m instead}}
\psfigRestoreAt

\newcommand{\Eq}[1]{Eq.~(\ref{#1})}
\newcommand{\Fig}[1]{Fig.~\ref{#1}}
\newcommand{\deffig}[4]{
  \begin{figure}[tbh]
    \begin{center}
      \null\ \\
      \null\
      \psfig{figure=#2,width=#3}
    \end{center}
    \caption[*]{#4}
    \label{#1}
  \end{figure}
}

\def\runtitle{
Zero-Temperature Critical Phenomena in 2D Spin Glasses
}
\def\runauthor{Naoki {\sc Kawashima} and Takayuki {\sc Aoki}}

\title
{
Zero-Temperature Critical Phenomena\\
in Two-Dimensional Spin Glasses
}

\author
{ 
Naoki {\sc Kawashima}\footnote{E-mail: nao@phys.metro-u.ac.jp}
and 
Takayuki {\sc Aoki}$^{1,}$\footnote{E-mail: t-aoki@nmit.tmg.nec.co.jp}
}

\inst
{
Department of Physics, Tokyo Metropolitan University,
Minamiohsawa 1-1, Hachiohji, Tokyo 192-0397, Japan\\
$^1$ 2nd System Engineering Department, NEC Microcomputer Technology, Ltd.,
Tsukagoshi 3-484, Saiwai, Kawasaki, Kanagawa 210-8511, Japan
}

\recdate
{
\today
}

\abst
{
Recent developments in study of two-dimensional spin glass models are 
reviewed in light of fractal nature of droplets at zero-temperature.
Also presented are some new results including a new estimate of 
the stiffness exponent using a boundary condition different from 
conventional ones. 
}

\kword
{
critical phenomena, spin glass, disordered system, droplet picture
}

\begin{document}
\sloppy
\maketitle

\section{Introduction}
One of most important unresolved issues concerning disordered
spin systems is whether the low-temperature phase of spin glass
systems is described by the mean-field picture~\cite{MeanField}
or rather by the droplet picture~\cite{FisherHuse}.
For three dimensions, the issue has not yet been settled
in spite of a lot of arguments and numerical works\cite{Arguments}.
On the other hand, most researchers believe that the spin glass model
in two dimensions
with a continuous bond distribution is disordered (paramagnetic) at
any finite temperature and is critical right at the zero temperature.
In this case, the critical phenomena near the zero temperature
have been understood mainly by the droplet picture~\cite{FisherHuseII} 
and other similar arguments such as the domain wall renormalization 
group argument~\cite{BrayMoore1984}.
The droplet picture~\cite{FisherHuseII} can also explain fairly well the 
experimental results on the two-dimensional spin glass systems~\cite{SGFilms}.
According to the standard droplet picture, the characteristic length scale 
at each temperature, i.e., the correlation length, 
is directly comparable to the typical size of an activated droplet.
Then, it is natural to expect that the various length scales
associated with the same energy (temperature) scale should be essentially
the same except for some constant prefactor regardless of 
difference in the definitions.
For example, the correlation length at temperature $T$ should be
the same as the typical droplet size whose excitation energy is $k_BT$.
This yields the scaling relation
$$
  -\theta_D = y_t
$$
where $y_t \equiv 1/\nu$ is the thermal scaling exponent,
and $\theta_D$ is the droplet exponent that characterizes
the size dependence of the droplet excitation energy.
Another example is the stiffness exponent $\theta_S$ that relates the size 
of the whole system to the excitation energy of the domain wall 
which is induced by some appropriately chosen
boundary condition.
The excitation energy is called domain wall energy or ``stiffness''
and hence the name of the exponent.
One, then, might expect that $y_t$, $-\theta_D$ and $-\theta_S$ are
equal to each other.
In fact, in the domain wall renormalization 
argument~\cite{BrayMoore1984,McMillan1984a},
it was concluded that $y_t = -\theta_S$.
However, we see in what follows that a number of numerical calculations,
such as the direct calculation of the magnetization as a function of
magnetic field~\cite{KawashimaS1992} and the recent direct measurement
of droplet excitation energies~\cite{Kawashima1999},
indicate that $\theta_S$ is truly different from the other two,
whereas the estimates of $y_t$ and $\theta_D$ are 
at least close to each other.

In this paper, we concentrate our attention on the Edwards-Anderson (EA) models
described by the following Hamiltonian,
$$
  {\cal H} = - \sum_{(ij)} J_{ij} S_i S_j,
$$
where $J_{ij}$ is a quenched random variable with some given bond distribution.
It is a general belief that models described by this Hamiltonian
can be classified into two classes according to the type of 
the distribution of $J_{ij}$.
One is the class of continuous distributions.
The Gaussian bond distribution is the most often 
studied example in this category.
Models in this category do not have non-trivial degeneracy
in the ground states.
The other class is that of discontinuous distributions to which
the $\pm J$ model belongs.
Models in this category in general have non-trivial degeneracy and 
the entropy per spin approaches to a finite value in the 
zero-temperature limit.
In other words, the model is not completely ``frozen'' 
even at zero-temperature, 
making the two-point correlation function decaying to zero.
As a result, they are characterized by non-zero values of the exponent $\eta$.

In section 2, we consider the model with continuous bond distribution.
We review some of previous numerical calculations for models in this 
category focusing on the three exponents, $\theta_S$, $y_t$ and $\theta_D$.
We also present some new results on $\theta_S$.
In section 3, we review briefly recent results on the discontinuous 
distributions.

\section{Models with Continuous Bond Distributions}

\subsection{Domain-wall renormalization-group argument and stiffness exponent}

\deffig{fg:Samples}{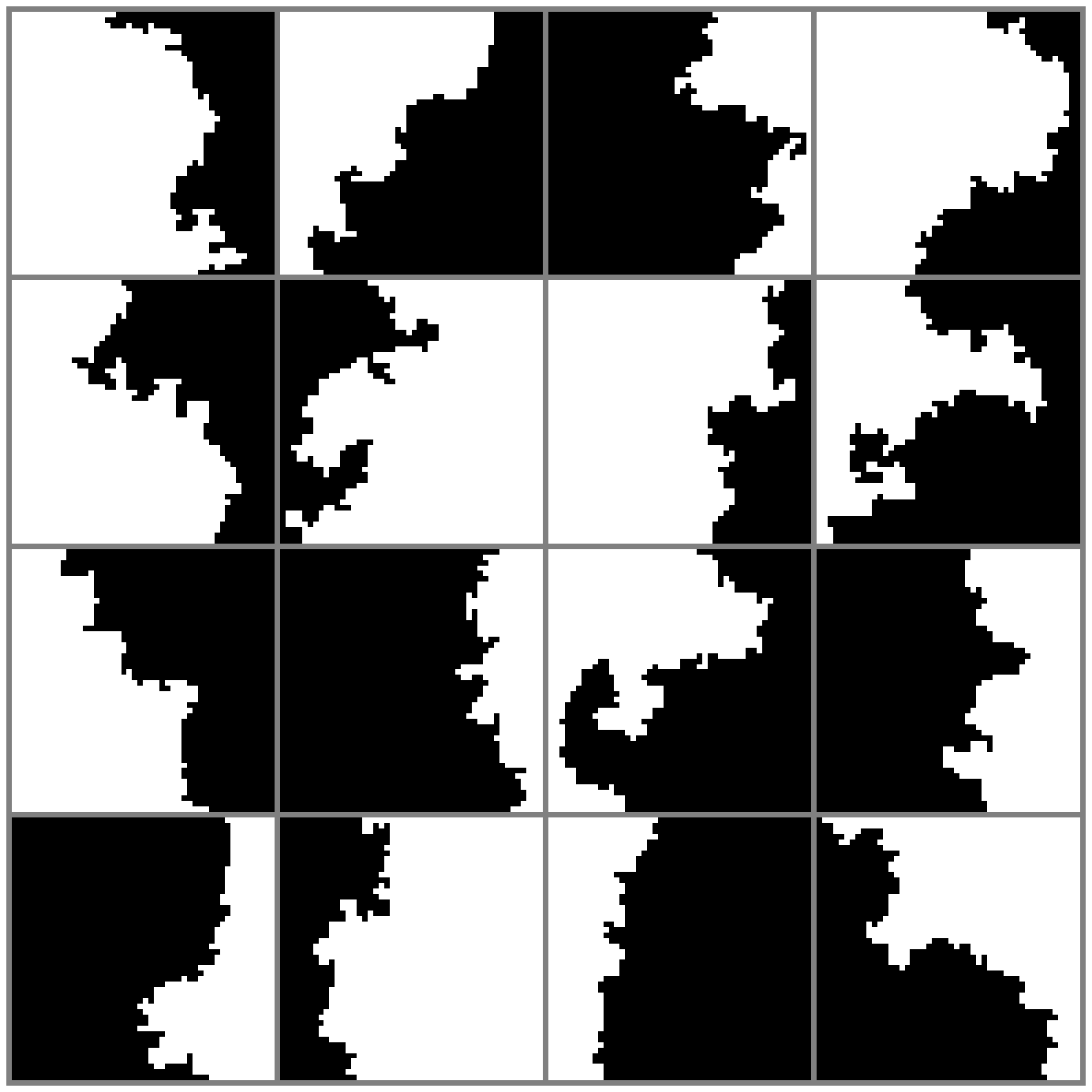}{70mm}
{The domain walls in 16 randomly chosen samples with $L=48$.}
It was argued~\cite{BrayMoore1984} that the excitation energy $E_W(L)$ 
of a domain wall across a system of size $L$ can be regarded as the
renormalized coupling constant after the system is renormalized up to 
scale $L$.
It follows that if we define the scaling exponent $\theta_S$ by
$$
  E_W(L) \propto L^{\theta_S},
$$
$\theta_S$ is expected to equal $y_t$.
Also, in the standard droplet theory for spin glasses~\cite{FisherHuse},
$E_W(L)$ is considered to be of the same order as the excitation 
energy of a droplet of scale $L$.
If these interpretations are correct, we can estimate the thermal
exponent $y_t$ or the droplet exponent $\theta_D$
simply by measuring the domain wall energy for systems of various sizes
instead of computing derivatives of the free energy or the droplet 
excitation energy.
In fact, it is technically much easier to compute the domain wall 
energies than droplet excitation energies.
This is one of the reasons why in the early stage of the study on spin glasses 
in two dimensions, many works were devoted to computation of the
domain wall energy. 
We can see this in Table I.
In Table I, some of previous numerical works for estimating one or two of
the three scaling exponents, $y_t$, $\theta_S$ and $\theta_D$,
for models with continuous, mostly Gaussian,
bond distributions in two dimensions are listed in chronological order.

\begin{fulltable}
\caption{
Some of previous estimates of scaling exponents for the two dimensional
EA model with the continuous bond distribution. 
The bond distribution is Gaussian for all the works listed here except 
for two items with footnotes.
$T_{\rm min}$ is the lowest temperature at which the computation for the largest system was performed.
In the last column, TM, MC, HO, EO and FF stand for the numerical transfer method (TM),
the Monte Carlo simulation (MC), a heuristic optimization (HO),
an exact optimization (EO) and mapping to a free fermion problem (FF), respectively
}
\label{tb:Exponents}
\begin{fulltabular}[t]{lllllll} \hline
\makebox[35mm][l]{Authors} & \makebox[15mm][l]{$-\theta_S$} & \makebox[15mm][l]{$y_t$} &
\makebox[15mm][l]{$-\theta_D$} & \makebox[20mm][l]{Size and Boundary Condition} & $T_{\rm min}$ & Method \\ 
\hline
Cheung \& McMillan~\cite{CheungM1983}     & 0.34(3)$^{[1]}$  & & & 11 (periodic) $\times$ $\infty$        & $\ge 0.15$ & TM \\
McMillan~\cite{McMillan1984a}             & 0.281(5)  & & & 8 (periodic) $\times$ 8 (p.t.$^{[2]}$)        & $\ge 0.3$  & TM \\
Bray \& Moore~\cite{BrayMoore1984}        & 0.291(2)  & & & 13 (randomly fixed) $\times$ 12 (periodic)  & 0 & TM \\
McMillan~\cite{McMillan1984b}             & 0.306(15) & & & 8 (periodic) $\times$ 8 (periodic)          & 0 & HO \\
Huse \& Morgenstern~\cite{HuseM1985}$^{[3]}$  & 0.24(3)$^{[1]}$ & & & 8 (periodic) $\times$ $\infty$              & 0 & TM \\
Cieplak \& Banavar~\cite{CieplakB1990}    & 0.31(2)   & & & 10 (randomly fixed) $\times$ 11 (periodic)  & 0 & TM \\
Kawashima \& Suzuki~\cite{KawashimaS1992} & & 0.476(5)  & & 20 (periodic) $\times$ 20 (periodic)        & 0 & HO \\
Kawashima et al.~\cite{KawashimaHS1992}   & & 0.48(1)   & & 16 (periodic) $\times$ 16 (free)            & $\ge 0.1$ & TM \\
Liang~\cite{Liang1992}                    & & 0.50(5)   & & 128 (periodic) $\times$ 128 (periodic)      & $\ge 0.4$ & MC \\
Rieger et al.~\cite{Rieger1997}           & 0.281(2)  & & & 30 (periodic) $\times$ 30 (periodic)        & 0 & EO \\
Rieger et al.~\cite{Rieger1997}           & & 0.48(1)   & & 60 (periodic) $\times$ 60 (periodic)        & 0 & EO \\
Nifle \& Young~\cite{Nifle1997}           & & 0.55(7)   & & 20 (periodic) $\times$ 20 (periodic)        & $\ge 0.8$ & MC \\
Huse and Ko~\cite{HuseK1997}$^{[4]}$      & & 0.37$^{[5]}$ & & 40 (p.t.) $\times$ 40 (p.t.)                & ---       & FF \\
Matsubara et al.~\cite{Matsubara1998}     & 0.2       & & & 21 (free) $\times$ 20 (periodic)            & 0         & MC \\
Kawashima~\cite{Kawashima1999}            & & & 0.47(5)   & 49 (free) $\times$ 49 (free)                & 0         & HO \\
present                                   & 0.290(10) & & & 48 (free) $\times$ 48 (free)                & 0         & EO \\
\hline
\end{fulltabular}
{\scriptsize
(1) ... The estimate of $1/\nu_{\parallel}$. (See text)\\
(2) ... Periodic tiling. (See text)\\
(3) ... The exponential bond distribution as well as the Gaussian distribution was used.\\
(4) ... The distribution used was symmetric, consists of two continuous parts, and has a vanishing weight around $J\sim 0$.\\
(5) ... This may not be considered as an estimate of $y_t$. (See text)\\
}
\end{fulltable}
The first estimate of the stiffness exponent was carried out through
the numerical transfer matrix method by McMillan~\cite{McMillan1984a}.
He applied the periodic or anti-periodic boundary condition 
in one direction while
replicating the same $L\times L$ system periodically in the other direction
to obtain a long strip, thereby mimicing the periodic boundary condition.
The width of the widest strip was $L=8$.
By switching from the periodic boundary condition to the anti-periodic one,
a domain wall is created across the strip.
The excitation energy of such domain walls at zero-temperature was estimated 
through an extrapolation of the domain wall free energy at finite temperatures
to zero temperature.
The resulting estimate of the stiffness exponent was
\begin{equation}
  -\theta_S = 0.281(5). \label{eq:McMillanEstimate}
\end{equation}

Bray and Moore\cite{BrayMoore1984} used the zero-temperature transfer matrix 
method in order to directly calculate the domain wall energy 
for slightly larger $L$'s than McMillan's previous calculation.
Instead of using periodic or antiperiodic boundary condition for creating 
domain walls,
they dealt with $L\times (L+1)$ systems with periodic boundary condition
in the direction of $L$ and randomly fixed boundary condition in the
direction of $L+1$.
Domain walls were created by simultaneous inversion of all the fixed spins 
on one of two boundaries perpendicular to the direction of $L+1$.
They obtained an estimate of the stiffness exponent only slightly smaller
than \Eq{eq:McMillanEstimate}.

McMillan\cite{McMillan1984b} obtained another estimate of the stiffness
exponent based on zero-temperature stiffness calculation through a heuristic 
optimization. An ordinary Monte Carlo method at a finite temperature was used
for generating many initial spin configurations. 
By quenching these spin configurations, approximate solutions were obtained.
If the number of trials is large enough, i.e., if one generates
sufficiently many approximate solutions, one can expect that the solution
with the lowest energy among them coincides the true ground state 
with fairly large probability.
In this way, $L\times L$ systems with periodic boundary condition were
examined up to $L=8$.
The resulting estimate turned out to be consistent with the above two estimates.

Rieger et al.~\cite{Rieger1997} performed a similar calculations
on a much larger scale with periodic boundary condition in both the directions.
They dealt with systems up to $L=30$ using an exact optimization method based on
the branch-and-cut algorithm~\cite{Groetschel1985}.
They obtained a value in good agreement with \Eq{eq:McMillanEstimate}.

Matsubara et al.~\cite{Matsubara1998} argued that the value of the
stiffness exponent depends on the boundary condition.
They performed a Monte Carlo simulation on systems of $L\times (L+1)$
up to $L=22$.
For the first system, the periodic and the free boundary condition 
was applied to the direction of $L$ and $L+1$, respectively.
Then, for the second system, spins on one of the two boundaries 
perpendicular to the direction of $L+1$
are fixed as they are in the ground state of the first system 
while those on the other boundary are fixed opposite to the first system.
The boundary condition in the other direction, i.e., the direction of $L$,
remains to be periodic.
They obtained an estimate of the stiffness exponent significantly larger
than previous estimates.
They argued that the stiffness tends to be estimated smaller when 
periodic boundary condition is imposed because it introduces an additional
tension into a system, and this was why they used free boundary condition 
for the first system.

\deffig{fg:Energy}{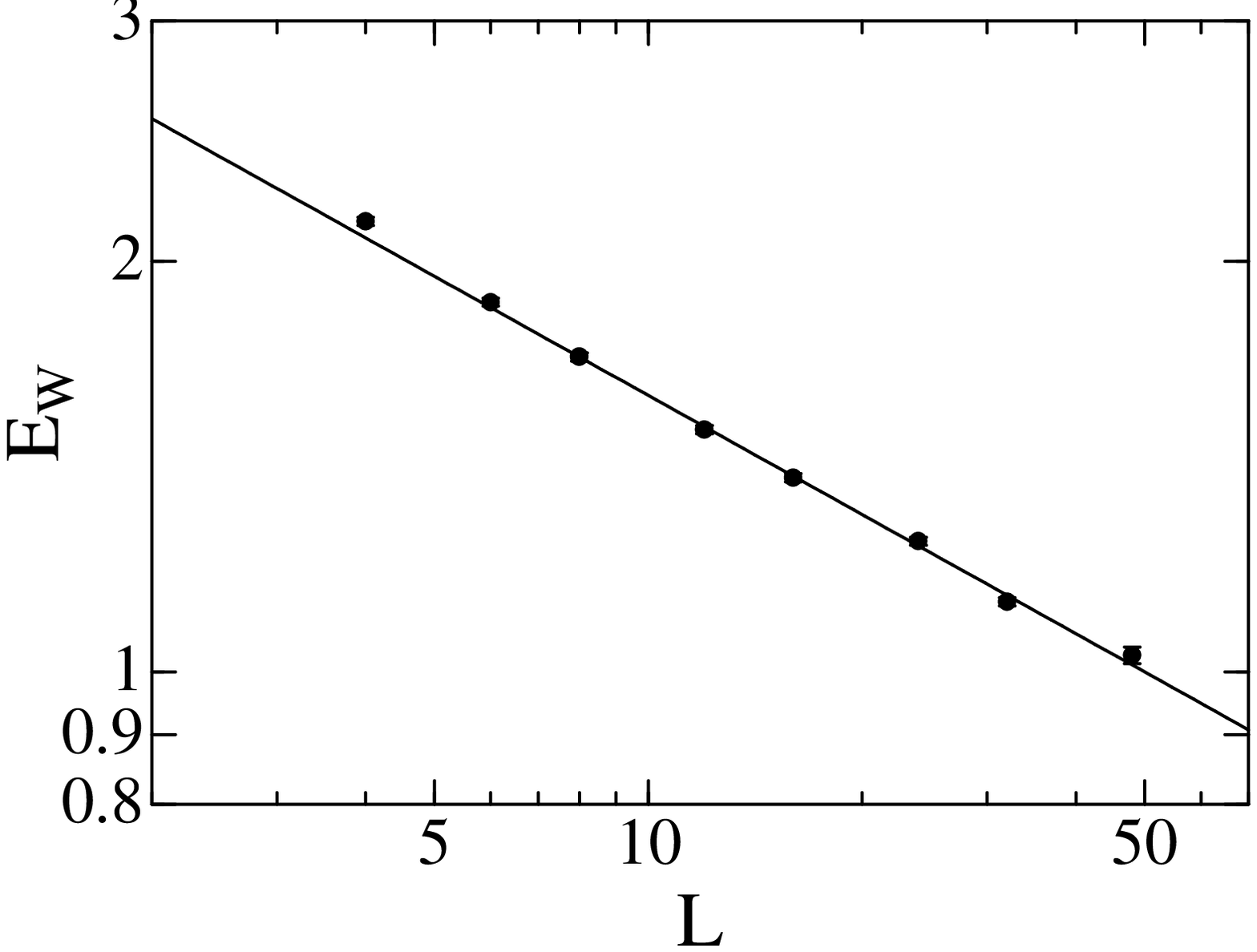}{80mm}
{The domain wall energy. 
The typical magnitude of statistical errors is about a quarter of the symbol size
except for the error bar of $L=48$ being slightly larger than the symbol size.}
In order to check if the stiffness exponent depends on the boundary condition,
we perform a numerical calculation of domain wall energy applying free 
boundary condition in every direction for the first system.
For the second system we apply a fixed boundary condition in one direction
just as Matsubara et al.\ did, i.e.,
we fix spins on one boundary as they are in the first system
and those on the other boundary opposite.
We carry out our calculation for $L\times L$ systems up to to $L=48$
using an exact optimization method~\cite{BarahonaMRU1982}.
In Fig.~\ref{fg:Samples}, some randomly chosen examples of the domain walls 
are illustrated.

In Fig.~\ref{fg:Energy}, the domain wall energy is plotted against the system
size in logarithmic scale.
The linearity is rather good and the slope is estimated as
$$
  -\theta_S = 0.290(10)
$$
in agreement with other previous results.
Here, the cited error has been estimated manually after doing some
trial-and-errors in the plot.
This error estimate turns out relatively conservative compared to previous estimates.
In fact, we should be conservative in the present case because
the exponent $\theta_S$ is small in absolute value which makes
the corrections to scaling relatively large.
For example, if we plot the excitation energy against $L+a$ instead of $L$, 
where $a$ is some constant term of order of unity,
the estimate of $\theta_S$ would change significantly.
This sort of correction due to a constant term added to the system size 
exists, in principle,
and should be taken into account in order to make the estimate reliable.
We suppose this is one of reasons for the discrepancy between
our estimate and Matsubara et al.'s estimate.

We also investigate geometrical properties of domain walls.
In Fig.\ref{fg:Perimeter}, averaged length along the perimeter $P$
and the roughness $R$ of droplets are plotted against $L$.
Here the roughness is the difference in the $x$-coordinates
of the left-most site and the right-most site on the domain wall.
\deffig{fg:Perimeter}{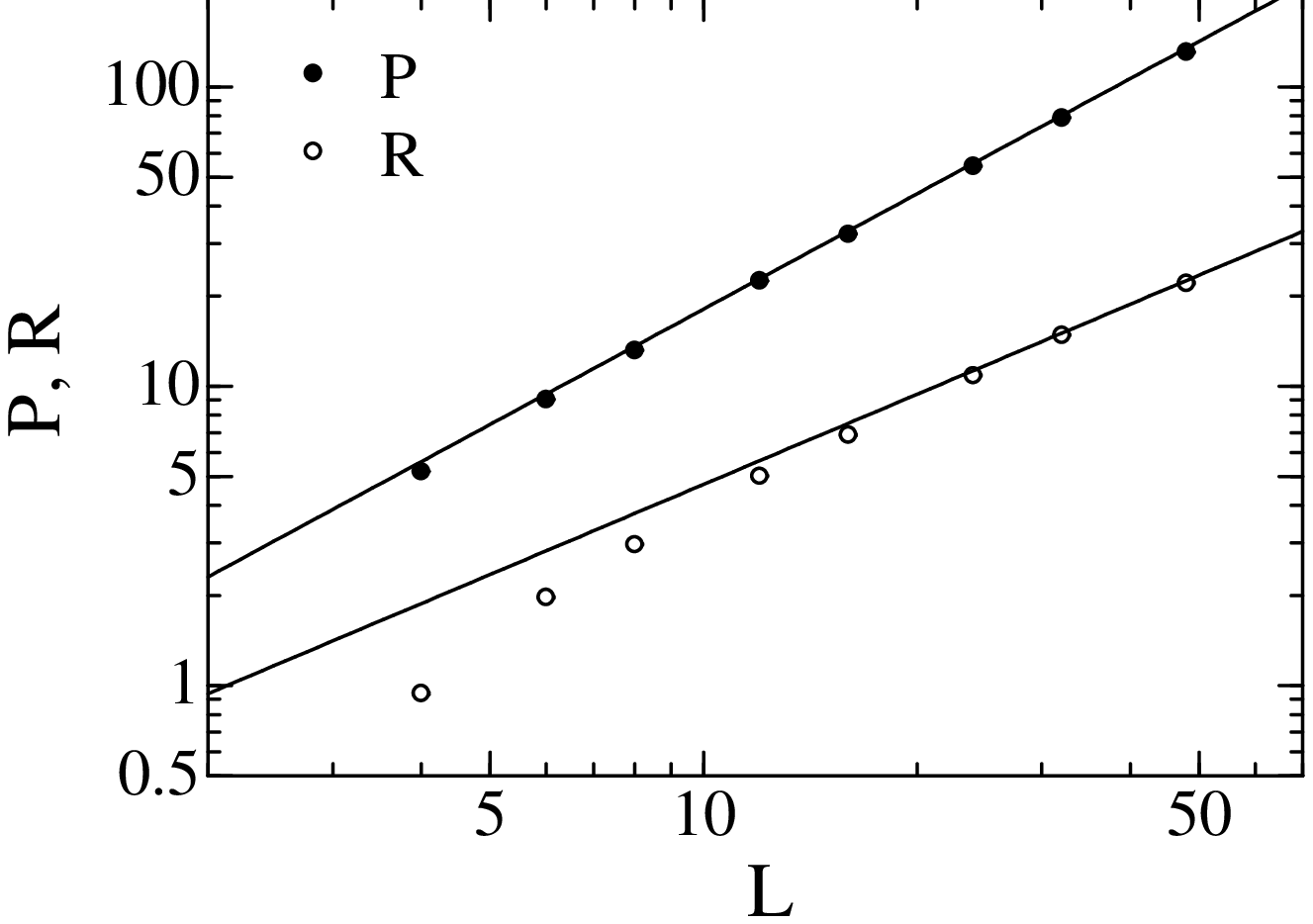}{85mm}
{The averaged perimeter $P$ and roughness $R$ of domain walls.}
The obtained data for the perimeter can be fit well by the following 
scaling form
$$
  P \propto L^{D_W} \quad \mbox{where}\quad D_W = 1.28(2).
$$
The estimate of $D_W$ agrees very well with the previous estimate
$D_W = 1.34(10)$\cite{Rieger1997}
for systems with the periodic boundary condition.
On the other hand, the data for the roughness in the logarithmic scale
clearly shows a systematic decrease in the slope converging to
unity around $L\sim 32$, which indicates a presence of 
a strong correction to scaling.
However, this non-linearity can be easily accounted for by
adding a constant to $L$ as discussed above.
The resulting estimate is consistent with 
$$
  R \propto L^{\zeta} \quad\mbox{where}\quad \zeta = 1.
$$

\subsection{Zero-temperature finite size scaling and the thermal exponent $y_t$}
As is discussed above, in the droplet argument or the domain-wall 
renormalization-group argument, we have only one characteristic length scale 
at each temperature.
It was also concluded, based on these arguments, that all the exponents that 
characterize the scaling properties of various static quantities can be 
derived from the droplet exponent or the stiffness exponent.
It should be noted, however, that we can derive this ``one-parameter-scaling''
without any special pictures or assumptions.
Instead, we simply assume the standard finite size scaling form for the singular 
part of logarithm of the partition function,
\begin{equation}
  \log Z_{\rm sing}(T,H,L) = f(TL^{y_t},HL^{y_h}),
  \label{eq:FSS}
\end{equation}
where $f$ is a scaling function and
$T,H,L$ are the temperature, the uniform magnetic field and
the system size, respectively.
From this form, we obtain for the magnetization the following expression,
$$
  M(T,H,L) = L^{y_h - y_t} \tilde m(TL^{y_t},HL^{y_h}).
$$
Since the model does not have a degeneracy in the ground state 
except the trivial one with respect to simultaneous inversion of
all spins, the magnetization in the limit of $H\to +0$
should be proportional to $L^{d/2}$.
This yields a scaling relation
$$
  y_h = y_t + \frac{d}{2}.
$$
Therefore we have
\begin{equation}
  M(T,H,L) = L^{d/2} \tilde m(TL^{y_t},HL^{y_t + d/2}). \label{eq:Magnetization}
\end{equation}
Then, it is easy to see that the zero-temperature
linear susceptibility at $H=0$ has the asymptotic form
\begin{equation}
  \chi(T=0,H=0,L) \propto L^{y_t}. \label{eq:Susceptibility}
\end{equation}
Similarly, the magnetization at zero-temperature has the form
$$
  m(T=0,H,L=\infty) \propto H^{1/\delta} \quad\mbox{where}\quad \delta=1 + y_t.
$$

There were some numerical attempts on computing thermodynamic 
quantities even before McMillan's calculation of stiffness.
Such measurements were done by Monte Carlo simulations~\cite{KinzelB}.
In retrospect, their estimates were affected by insufficient equilibration
in the low temperature region.
The first direct numerical calculation of a thermodynamic quantity
after researchers started taking into account such equilibration problems
was carried out through an optimization method 
by Kawashima and Suzuki~\cite{KawashimaS1992}.
We obtained ground states of the systems with Gaussian bond distribution,
thereby measuring the magnetization at a small but finite magnetic field.
In particular the zero-temperature linear susceptibility for various
system sizes up to $L=20$ was measured.
By matching the resulting estimate with \Eq{eq:Susceptibility}, 
the thermal exponent was estimated as
\begin{equation}
  y_t = 0.476(5).\label{eq:FirstEstimate}
\end{equation}
Surprisingly, the value did not agree with the stiffness exponent 
which had been estimated rather accurately.

From Eq.(\ref{eq:FSS}) the scaling form for the
spin glass susceptibility can be also derived. Namely,
\begin{eqnarray}
  \chi_{\rm sg} &\equiv& 
  \frac1{L^d}\sum_{ij} [\langle
  S_i S_j \rangle^2]
  \sim - \frac{1}{3L^d}\frac{\partial^4}{\partial (\beta H)^4} 
  \log Z_{\rm sing.} \nonumber \\
  & \sim & L^d \tilde \chi_{\rm sg}(TL^{y_t},HL^{y_t + d/2})
  \label{eq:FSSXSG},
\end{eqnarray}
where $[\cdots]$ stands for the bond configuration average whereas
$\langle \cdots \rangle$ the thermal average.
The spin glass susceptibility was computed numerically through
the transfer matrix method~\cite{KawashimaHS1992} for system
sizes up to $L=16$ at various temperatures.
From the best fit to the form Eq.(\ref{eq:FSSXSG}) with setting $H=0$,
the exponent $y_t$ was estimated as
$$
  y_t = 0.48(1),
$$
which agrees with the previous result obtained through the computation
of the magnetization.

Liang~\cite{Liang1992} also computed $\chi_{SG}$ for larger systems 
by a cluster Monter Carlo method.
It was at $T=0.4$ that the simulation of the largest
size ($L=128$) was performed.
The estimate of the thermal exponent was $y_t = 0.50(5)$
reconfirming the above-mentioned previous results.
Assuming that the temperature dependence of the correlation length
obtained by Cheung and McMillan~\cite{CheungM1983} is correct down to this
temperature, we can roughly estimate the correlation length at $T=0.4$
to be a few tens of lattice constants.
Therefore, Liang's calculation can be considered complementary to 
the above two previous calculations, 
in that Liang's calculation was performed at temperatures higher
than the cross-over temperature
whereas for the two calculations mentioned above were carried out
at temperatures lower than that.

The computation of the zero-temperature magnetization was also redone 
for larger systems.
Rieger et al.~\cite{Rieger1997} used an exact optimization method
for computing the magnetization in a magnetic field
for systems with periodic boundary condition.
Matching the data for systems up to $L=50$ to \Eq{eq:Magnetization}, 
they obtained an estimate of $y_t$ essentially identical to the previous 
one \Eq{eq:FirstEstimate}.

%
Given these results on the thermal exponent
together with the above mentioned estimates of stiffness exponent,
it is now rather unlikely that the discrepancy between them
is only a numerical artifact.
If they are truly different as the numerical estimates suggest,
a simple domain-wall renormalization-group argument~\cite{BrayMoore1984}
and an over-simplified droplet picture.

\subsection{Droplet argument and droplet exponent}
Then, how can we elaborate the drolet picture in order to explain
the discrepancy?
A recent work\cite{Kawashima1999} on droplets may shed light on this question
though the answer has not yet been obtained.
\deffig{fg:DropletSamples}{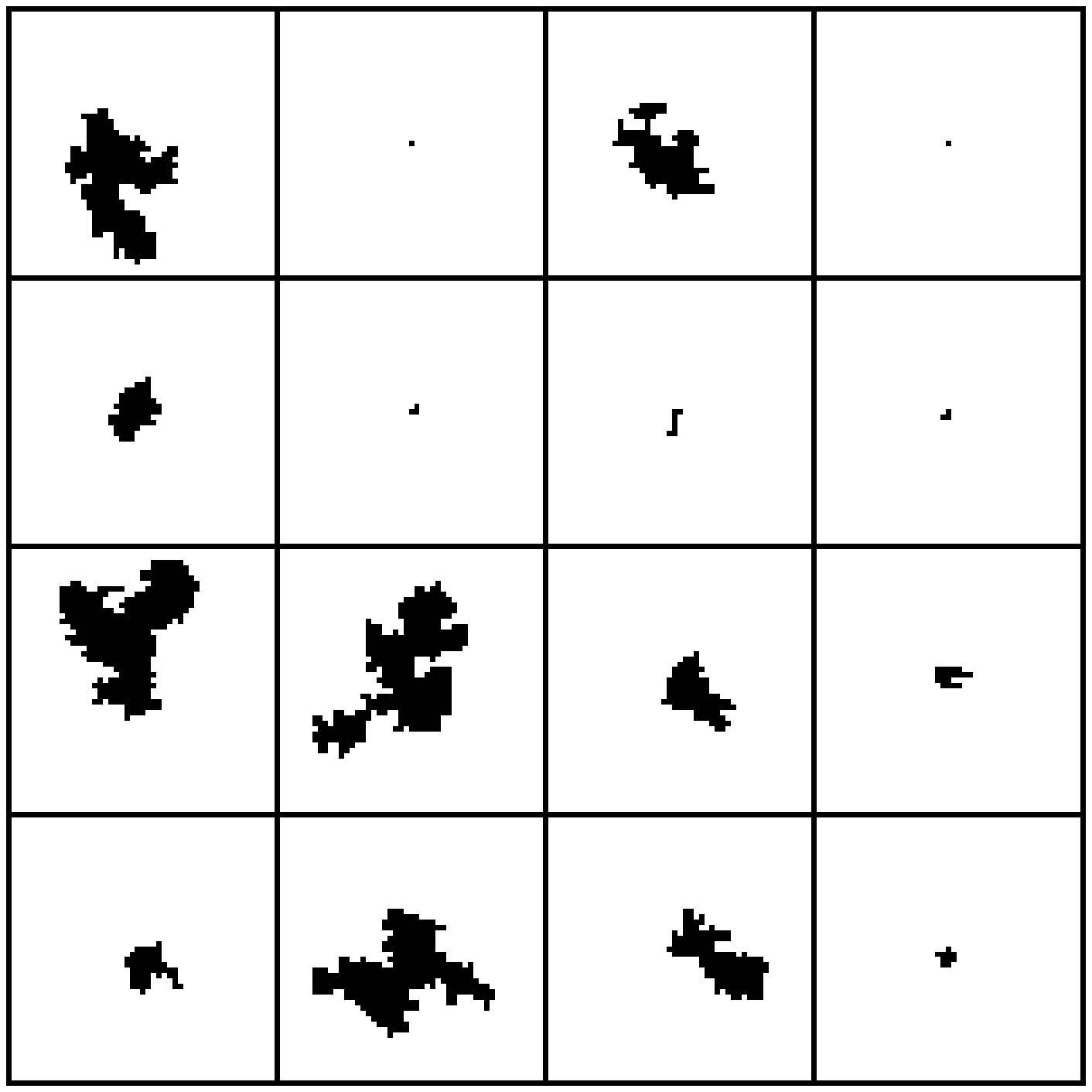}{70mm}
{Droplets of 16 randomly chosen samples.}
``Droplet'' is a well-known and useful concept in the study of
critical phenomena\cite{MFisher}. 

The droplet argument for spin glasses\cite{FisherHuse}
based on this concept is one of important working 
hypotheses in the field.
Droplets are collective excitations 
from some ordered state below the transition temperature.
A naive interpretation of the droplet theory leads to the conclusion 
$-\theta_S = -\theta_D = y_t$.
However, because of technical difficulty mentioned below,
no direct observations of the droplets was carried out until quite recently.
In addition, as we have seen above, at least the simplest version of
the argument does not work in the present case.
Therefore, the equivalence between $y_t$ and $-\theta_D$ 
should be re-examined carefully.

Fisher and Huse defined\cite{FisherHuse}
a droplet of scale $\lambda$ including a given site $i$ as
a cluster of spins (with $i$ among them) with the smallest 
excitation energy that contains more than $\lambda^d$ and 
less than $(2\lambda)^d$ spins.
The basis of the argument is the following scaling 
form that describes the excitation-energy distribution of droplets of
scale $\lambda$:
\begin{equation}
  P_{\lambda}(E_{\lambda}) = 
  \frac{1}{\Upsilon {\lambda}^{\theta_D}}
  \tilde P\left(\frac{E_{\lambda}}{\Upsilon {\lambda}^{\theta_D}}\right)
  \label{eq:DropletEnergyScaling}
\end{equation}
where $\Upsilon$ is some constant, $\theta_D$ is the droplet exponent, 
and $\tilde P(x)$ is the scaling function which is continuous and 
non-vanishing at $x=0$.
Since the droplet ``size'', $\lambda$, is defined to be 
proportional to (volume)$^{1/d}$,
it is not necessarily a spanning length of droplets 
because the volume of a droplet may in general 
have a non-trivial fractal dimension.
While it was argued \cite{FisherHuse} that the droplets must be compact
if $\theta_D$ is positive, no extensive discussion had been devoted 
to the geometrical properties of droplets in the present two-dimensional case,
where $\theta_D$ is predicted to be negative,
until quite recently.
We demonstrated~\cite{Kawashima1999} that 
a typical large droplet in two dimensions occupies only an 
infinitesimal fraction of the volume of the minimal box that can contain 
the droplet.
It follows that for a given droplet we have at least two essentially
different length scales, $\lambda$ and the spanning length.
We refer hereafter to the fractal dimension of the droplet volume as $D$.
Then, we have the relation 
$$
  V = \lambda^d \propto l^D
$$
where $l$ is the spanning length of the droplet.

Since the above-mentioned definition of droplets by Fisher and Huse
is inconvenient from computational point of view,
we adopted the following alternative definition.
First we consider an $L\times L$ system with free boundary condition.
A droplet of scale $L$ is then defined to be the cluster of spins
that has the smallest excitation energy among those which
contain the central spin and contains no spins on the boundary.

If we assume the conventional droplet argument,
the central spin is surrounded by droplets (in the original definition) 
of various scales.
Considering the fact that larger droplets tend to have
smaller excitation energy,
one may expect that the spanning length of the droplet 
with the new definition would be proportional to $L$.
We here define $\theta_D'$ by the following $L$ dependence of 
the average excitation energy of the droplets in our definition,
$$
  [E_D(L)] \propto L^{\theta_D'}.
$$
Comparing this to Fisher and Huse's definition of the droplet,
we obtain
\begin{equation}
  \theta_D = \frac{d}{D}\theta_D', \label{eq:Relationship}
\end{equation}
because $L$ is proportional to the spanning length of the droplet.

In order to observe droplets with the new definition,
we first computed the ground state with free boundary condition,
and took it as the reference spin configuration.
Then we computed the ground state with the constraint that
the spins on the boundary are to be fixed as they are in the reference state 
while the central spin is to be fixed opposite, 
thereby forcing a cluster of spins including the central spin to flip.
For a system with the free boundary condition, 
polynomial-time optimization algorithms are available 
whereas for the systems with constraints no such algorithm is known.
In fact, the two dimensional spin glass problem with general 
constraints has been proven to be NP hard\cite{BarahonaMRU1982}.
Therefore, we have employed the replica optimization\cite{KawashimaS1992},
which is a heuristic optimization algorithm based on the idea of
renormalization group.
The details of the algorithm are described elsewhere
\cite{KawashimaS1992,KawashimaFuture}.

Observed droplets varies in size and shape.
Some of them contains only one spin while spanning length of some others 
turned out to be comparable to the system size itself (See \Fig{fg:DropletSamples}).
This reflects the fact that $|\theta_D|$ is small
and the excitation energy does not very strongly depend on the droplet size,
yielding a non-negligible probability for a small droplets being chosen.
In the larger droplets, on the other hand, many handles and overhangs
can be observed, which already suggests the fractal nature of the droplets.
The averaged spanning length was found to be proportional to the system size
as we expected.

The averaged length, $P$, of the boundary of the droplet and
the averaged volume, $V$ were also measured.
For $P$, we obtained $ P \propto L^{D_s} $
with the surface fractal dimension
$$
  D_s = 1.10(2).
$$
For $V$, we estimated the fractal dimension as
\begin{equation}
  D = 1.80(2). \label{eq:D}
\end{equation}
The fractal dimension $D$ of droplets is certainly smaller than $d=2$.
Thus, we concluded that the droplets at the critical point $T=0$
have fractal nature in the volume as well as in the perimeter.

%
It was also found that the droplet excitation energy $E_D(L)$ 
have a broad distribution, similar to the size and the shape.
When rescaled with the average value $[E_D(L)]$,
histograms of excitation energies
for various system sizes fit on top of each other, 
showing the validity of the scaling form \Eq{eq:DropletEnergyScaling}.
In addition, we observe that the scaling function $\tilde P(X)$ 
has a non-vanishing value at $X=0$, satisfying a necessary condition
for the droplet argument to be valid.

%
When $\log E_D$ is plotted against $\log L$ the slope is estimated
to be $-0.42$.
From this value, together with two others obtained by
using other measures of the droplet size instead of $L$,
we obtained the estimate of the droplet exponent $\theta_D'$:
\begin{equation}
  -\theta_D' = 0.42 (4). \label{eq:DropletExponent}
\end{equation}
Assuming \Eq{eq:Relationship} and using the value of $D$ in \Eq{eq:D},
we obtained the following estimate of $\theta_D$,
\begin{equation}
  -\theta_D = 0.47(5)
\end{equation}
in good agreement with previous estimates of $y_t$ such as \Eq{eq:FirstEstimate}.

\subsection{Other related estimates of scaling exponents}
As far as we know, the first important attempt to estimate
$y_t$ was made through the computation of correlation length~\cite{CheungM1983}.
The numerical transfer matrix method was employed for strips of $L\times N$ where
$L\le 11$ and $N$ is so large that we can take it as infinite effectively.
The periodic boundary condition was imposed across the strip.
Correlation lengths $\xi_{\parallel}(T,L)$ in the longitudinal direction, i.e., 
the direction of $N$, were measured at various temperatures down to $T=0.15$.
They estimated the correlation length exponent as $\nu_{\parallel} = 2.96(22)$.
We should note here that this $\nu_{\parallel}$ is not necessarily equal to 
$\nu\equiv 1/y_t$ because of the very large aspect ratio of the systems.
As we have seen above, a number of numerical results suggest that domain walls 
and droplets of the same length scale correspond to different energy scales,
and it is the droplet excitation energy, not the domain wall energy,
that scales with the thermal exponent $y_t$.
On the other hand, in a long strip, it is the domain wall type excitations
that determines the correlation between two points seperated by a very long
distance in the longitudinal direction.
Therefore it is not very surprising that their estimate turned out to be
close to $\theta_S$, not $y_t$.

Another computation based on the measurement of correlation lengths was 
performed by Huse and Morgenstern\cite{HuseM1985} for models with the 
exponential bond distribution as well as those with the Gaussian distribution.
Again, the transfer matrix method was used.
The maximum width of the strip was $L=8$.
Assuming the phenomenological renormalization group relations,
$$
  \frac12 \xi_{\parallel}(T,2L) = \xi_{\parallel}(T',L)
$$
with
$$
  \frac{T'}{T} \approx 2^{-1/\nu_{\parallel}},
$$
they obtained the estimate $\nu_{\parallel} \sim 4.2 (5)$.

Huse and Ko~\cite{HuseK1997} carried out a unique estimation of an exponent.
They considered a system that consists of an infinite periodic repetition
of many $L\times L$ systems identical to each other.
Since the system size is infinite, it has a phase transition, presumably, 
of the Ising type.
Mapping the problem into a free fermion problem
the eigen values of its row-to-row transfer matrix were obtained
within a polynomial computational resources.
Transition temperatures, then, correspond to those at which the gap between
the largest and the second largest eigen values vanishes.
Assuming that the typical transition temperature is proportional
to $L^{-y_t}$ where $L$ is the size of the unit cell,
they obtained an estimate $y_t = 0.37$.
However, again because of the existence of multiple length scales,
it is not clear whether the exponent estimated in this way 
should be considered as $y_t$.

\section{Discontinuous Bond Distributions}
\deffig{fg:PhaseDiagram}{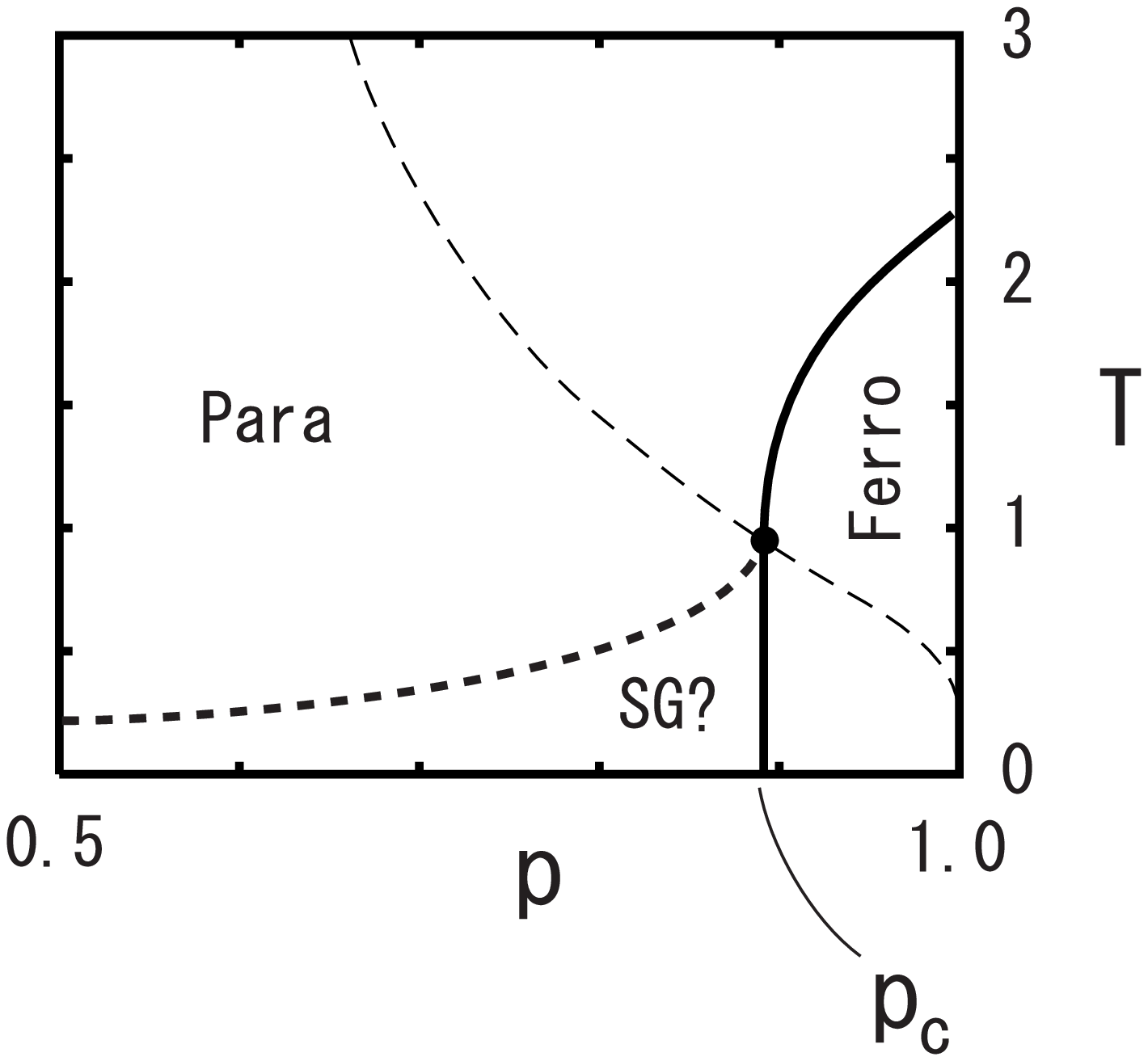}{70mm}
{The schematic phase diagram of $\pm J$ model on a square lattice.
The thin dashed line stands of the Nishimori line, not a phase boundary,
whereas the dotted line stands for a speculative phase boundary which
has not yet been confirmed.}
The critical behavior of the systems with discontinuous bond distributions is 
believed to be qualitatively different from the continuous distribution 
discussed in the last section because of the high degeneracy in the ground 
state.
In this case the critical properties are not as fully understood as 
in the case of continuous distributions.
What is missing is a theory corresponding to the droplet argument
for the case of continuous distributions.

The bond distribution in this class that is most often considered 
is the binary bond distribution where a coupling $J_{ij}$ takes on the 
value $+J$ and $-J$ with probability $p$ and $1-p$, respectively.
For this bond distribution, 
the possibility of a finite temperature phase transition 
even in the case of the symmetric distribution ($p=1/2$) has not 
been ruled out\cite{MatsubaraSS1997,KitataniS1999} as mentioned below.

\subsection{Intermediate Phase?}
Another problem that had been remaining to be solved until recently
concerns the $p-T$ phase diagram of the models with asymmetric 
binary bond distributions.
It had been suggested\cite{Barahona82,MaynardR,Ozeki90}
that some intermediate phase exists in the $p$-$T$ phase diagram
close to the vertical part of the phase boundary (See \Fig{fg:PhaseDiagram})
between the ferromegnetic phase and the other, presumably paramagnetic, phase.
The suggestion was made based on zero-temperature numerical calculations.
For the site-random model the evidence for the existence of 
an intermediate phase seemed to be even stronger than for the
bond-random model \cite{ShirakuraM95,OzekiN1995}.

We reinvestigated\cite{KawashimaR1997} this issue by
studying the domain wall energy at zero temperature via the
determination of exact ground states for large system sizes ($L\le 32$)
and large number of samples ($N_{\rm sample} \ge 32768$)
using a polynomial algorithm described 
by Barahona et al.\cite{Barahona82}.
We calculated the domain wall energy $E_W$ for various system sizes
with various ferromagnetic bond concentration $p$,
applying the periodic or the anti-periodic boundary condition in one
direction and the free boundary condition in the other.
The stiffness exponent $\theta_S$ and another exponent $\rho$ defined by
\begin{equation}
  [E_W] \propto L^{\rho}
\end{equation}
were estimated.
A positive value for $\rho$ signifies the stability of a ferromagnetic
ground state even in the presence of thermal fluctuations and therefore the
existence of the ferromagnetic long range order at finite temperature
\cite{BrayMoore1984}.
We defined $p_c^{(1)}$ and $p_c^{(2)}$ as the critical
concentrations of ferromagnetic bonds at which 
$[E_W]$ and $[E_W^2]^{1/2}$, respectively, change from increasing to decreasing
as functions of $L$.

We hypothesized the following finite size scaling forms for $E_W$ and $E_W^2$
\begin{eqnarray}
  [E_W] L^{\psi^{(1)}}
  & = & f_1( (p-p_c^{(1)}) L^{y_p^{(1)}} ), \label{eq:FSSI} \\
  {[E_W^2]}^{1/2} L^{\psi^{(2)}}
  & = & f_2( (p-p_c^{(2)}) L^{y_p^{(2)}} ). \label{eq:FSSII}
\end{eqnarray}
The parameters were estimated as follows.
\begin{eqnarray}
  & & p_c^{(1)} = 0.896(1),\ y_p^{(1)}=0.77(1),\ \psi^{(1)}=-0.19(2), 
  \label{eq:EstimateBondI} \\
  & & p_c^{(2)} = 0.894(2),\ y_p^{(2)} = 0.79(6),\ \psi^{(2)} = -0.16(4). 
  \label{eq:EstimateBondII}
\end{eqnarray}
The value of $p_c^{(1)}$ is consistent with most of previous estimates
such as 0.88(2) \cite{MorgensternB80}, 0.89(2) \cite{Barahona82} and 0.89(1)
\cite{ONishimori} while inconsistent with 0.885(1) \cite{Ozeki90}.
The estimate of $p_c^{(2)}$ is larger than but marginally consistent 
with all the previous estimates such as
0.86(2)  \cite{Sadiq1981},
0.85     \cite{Barahona82} and
0.870    \cite{Ozeki95b},
while it is clearly inconsistent with 0.854(2) \cite{Ozeki90}.
The coincidence between $p_c^{(1)}$ and $p_c^{(2)}$ suggests 
the absence of the intermediate phase.

It is also remarkable that not only $p_c^{(2)}$ but also $y_p^{(2)}$ and
$\psi^{(2)}$ agree with the corresponding values in \Eq{eq:EstimateBondI}
within the statistical errors.  While the agreement in $p_c$ already
suggests the absence of the intermediate phase, we consider the
agreement in the critical indices as another evidence for the absence
of the intermediate spin-glass phase, since otherwise it is hardly imaginable
that the first and the second moment of $E_W$ show the same
critical behavior at different values of $p_c$.

We also computed the stiffness for the site-random model.
We found that $p_c^{(1)}$ and $p_c^{(2)}$ for the site-random
model agreed with each other 
and that the first and the second moments were scaled with
the same scaling exponents as the bond-random model mentioned above,
though with larger corrections to scaling.

Prior to our work, Nishimori argued \cite{Nishimori1986} that 
the phase transition is of purely geometrical nature,
which leads to the strictly vertical phase boundary that seperates
the ferromagnetic phase from the paramagnetic (or glassy) phase in the
low temperature region.
If this is the case, the value of $p_c$ at $T=0$ must coincide
that of the multicritical point, i.e., the solid circle in the phase
diagram \Fig{fg:PhaseDiagram}.
The most recent calculation for the multicritical point was done
by Ozeki and Ito \cite{OzekiI1998}, who performed Monte Carlo simulations 
at finite temperatures along the Nishimori line to locate precisely 
the multicritical point in the phase diagram.
It is unsettling that previous estimates of $p_{\rm m.c.}$\cite{OzekiN1987,SinghA1996},
including Ozeki and Ito's estimate $p_{\rm m.c.} = 0.8872(8)$,
are slightly smaller than our estimate of $p_c$ beyond
the extent of errors although the difference is rather small.
The easiest explanation is that systematic errors 
might not be fully taken into account in the estimates, 
resulting in an underestimate of the errors.
Another possibility is that the size dependence,
\Eq{eq:FSSI} and \Eq{eq:FSSII},
assumed in the estimation of the exponents may not correctly describe the true behavior.
The last possibility is that the phase boundary may not be strictly 
vertical in spite of Nishimori's argument.
So far, we have not yet identified the reason for the discrepancy.

\subsection{Relation to the Percolation Transition}
We should notice the coincidence between the exponent $y_p^{(1)}$
(or $y_p^{(2)}$) in the above mentioned work\cite{KawashimaR1997}
with the scaling exponent $y_p \equiv 1/\nu_p$ where $\nu_p$ is
the correlation length exponent for the percolation tranition in two dimensions.
Related to this finding, in a work preceding our estimation of $y_p^{(1)}$ and $y_p^{(2)}$,
Singh and Adler\cite{SinghA1996} investigated the critical behavior
around the multicritical point,
and discussed its relation to the percolation transition.
They suggested that one of the two scaling axes at the multicritical point
is parallel to the Nishimori line whereas the other is parallel to the 
temperature axis.
They estimated the scaling exponent which corresponds to our $y_p^{(1)}$
and found that the value agrees with the percolation exponent.
Since, according to Nishimori's argument, the temperature plays 
no essential role along the vertical part of the phase boundary,
we could argue that the phase transitions across the boundary
are characterized by the same scaling exponent 
independent of the temperature,
although we are not sure if the argument should be valid
at the two end points as well.

\subsection{Finite-Temperature Phase Transition at $p=1/2$?}
In the case of the bond-random $\pm J$ model with $p=1/2$,
the majority of researchers tend to think that there is no phase transition
at a finite temperature.
This general belief is based on results of Monte Carlo simulations
\cite{BhattYoung,MorgensternB},
high-temperature series expansion\cite{SinghC1987},
and the estimates of the domain wall
energy \cite{BrayMoore1984,CieplakB1990,Ozeki90}. 
The data from Monte Carlo simulations
are not available at very low temperature,
e.g., below $T=0.4$ in Bhatt and Young's simulation \cite{BhattYoung}, 
which makes it difficult to exclude the possibility of the
transition at a temperature smaller than $0.4$.
Although the results of high-temperature series expansion also 
suggested the absense of the finite temperature phase transition,
it was not very conclusive.
As for the calculations of the domain wall energy,
the ``stiffness'' exponent turned out to be even smaller 
in absolute value than that for the continuous distributions.
Therefore, it was not very clear if $\theta_S$ is really negative.
For example, the data in Cieplak and Banavar's paper\cite{CieplakB1990} 
clearly shows a systematic positive curvature 
in the plot of the domain wall energy versus the system size.
In short, so far we have not obtained very convincing evidences 
supporting the general belief, i.e., the absense of the phase 
transition at a finite temperature in two dimensions.

On the other hand,
results of Monte Carlo simulation at lower temperatures
were reported\cite{ShirakuraM96} indicating a transition at
$T\simeq 0.24$ for the $\pm J$ model with $p=1/2$.
A finite temperature phase transition was suggested\cite{ShirakuraM97}
also for the model with another discrete distribution where
a bond variable takes on $J$ or $-aJ$ where $0 < a < 1$.
In addition, $\theta_S=0$ was suggested \cite{Ozeki90} based on 
estimates of the domain-wall energy. 
These results are consistent with a finite temperature phase transition 
for which the low-temperature phase is only marginally or weakly ordered, 
i.e., the two-point spin-spin correlation function decreases
algebraically as a function of the distance.

On this issue, we found\cite{KawashimaR1997} that 
the stiffness decreases systematically 
but it does so extremely slowly in the case of $p=1/2$.
The obtained data was consistent with the algebraic decrease
as a function of the system size with the stiffness exponent 
$\theta_S = -0.056(6)$.
However, we should note that the quoted error here 
only includes the estimated statistical error
that is obtained in the standard procedure of the method of least squares.
Because of the very weak dependency on the system size,
the systematic error is more important here than the statistical error.
Since the systematic error is difficult to estimate,
other scenarios, such as convergence to a finite constant\cite{Ozeki90} 
or logarithmic decay, are equally probable.
If either one of these scenarios is correct, 
we cannot determine based solely on calculations of the stiffness exponent
whether the long-range order persists at a low but finite temperature.
At the moment, what we can conclude is only that
the low-temperature phase is only weakly ordered 
even if the phase transition takes place at a finite temperature.

Very recently, Kitatani and Sinada\cite{KitataniS1999} performed 
an interesting numerical calculation of spin glass susceptibility 
and its Binder ratio.
Based on the invariant properties under the gauge transformation,
they constructed a new method for computing cumulants
of Edwards-Anderson order-parameter at any temperature using
a novel Monte Carlo method in the bond configuration space 
rather than in the spin configuration space.
Their results of the finite size scaling analysis,
appeared to suggest the existence of a phase transition around $T = 0.3$.
However, based on the data analysis with corrections to scaling,
they concluded that the scenario of the zero-temperature phase transition 
could explain their data as well.

\section{Concluding Remarks}
We have reviewed recent developments in study of spin
glass models in two dimensions.
It has become clear in the case of continuous bond distributions
that the critical behavior is dominated by fractal droplets 
in contrast to the compact droplets
in conventional droplet argument for higher dimensions.

The disagreement between the droplet exponent 
and the stiffness exponent remains unsettling.
At least, however, the present result suggests that
a domain wall of a $L\times L$ system may not necessarily be considered
as an object of scale $L$ because of the existence of different 
ways for measuring the size of the same object.
In other words, at present we do not know which way of measuring
we should use for comparing the size of a domain wall
with that of a droplet.

The models with discrete distributions are relatively poorly understood.
It is an important and challenging future problem
to determine whether a finite temperature phase transition
takes place or not by larger scale numerical calculations.

Although it is not yet clear whether fractal objects similar
to the present two-dimensional case can be
defined at the critical point in higher dimensions, 
it is certainly worth thinking about such a possibility.
The essential difference would be that $\eta$ is not in general zero in
higher dimensions whereas $\eta=0$ for continuous distributions 
making the theory simpler in the latter case.
Provided that the $\pm J$ model in two dimensions is critical 
at $T=0$, the same remark may apply to this discrete model
since $\eta$ is non-vanishing there, too.
Therefore, we may obtain some new perspective for both the finite-temperature
transition in three dimensions and the zero-temperature critical behavior
for discrete distributions in two dimensions
by generalizing the concept of zero-temperature fractal droplets
to the cases with non-vanishing $\eta$.

\section*{Acknowledgements}
This work is supported by Grant-in-Aid for Scientific Research Program
(No.11740232) from Mombusho, Japan.

\end{document}